\documentclass[a4paper,12pt, epsfig]{article}
\def\letter{0}\def\pr{0}
\pdfoutput=1 
\usepackage{epsfig}
\usepackage{epstopdf}
\usepackage{graphicx}
\usepackage{ifthen}
\usepackage{ulem}

\usepackage{amsmath}
\usepackage[psamsfonts]{amssymb}
\usepackage{euscript}

\usepackage{latexsym}
\usepackage[arrow,matrix,curve]{xy}

\usepackage[hypertexnames=false]{hyperref}
\hypersetup{
    colorlinks=true,
    linkcolor=blue,
    filecolor=magenta,   
    citecolor=black,   
    urlcolor=cyan,
    }

\newskip\humongous \humongous=0pt plus 1000pt minus 1000pt

\newif\ifdtup

\def\,{\hspace{-.1cm}}
\def\hsp{,\hspace{.7cm}}

\def\fc#1#2 {\frac{n}{q}#1\frac{n}{q}#2}

\newcommand{\vac}{\ensuremath{|0\rangle}}

\def\exp#1{\hbox{\rm exp}\left(#1\right)}

\renewcommand{\theequation}{\arabic{section}.\arabic{equation}}
\renewcommand{\(}{\begin{equation}}
\renewcommand{\)}{end{equation} \vspace{-.05in}\linebreak}

\newcounter{saveeqn}
\newcounter{savealpheqn}

\newcommand{\alpheqn}{\setcounter{saveeqn}{\value{equation}}%
  \stepcounter{saveeqn}\setcounter{equation}{0}%
  \renewcommand{\theequation}{\mbox{\arabic{section}.\arabic{saveeqn}
\alph{equation}}}
  \renewcommand{\)}{\end{equation}}}
\def\part#1{\frac{\partial}{\partial{#1}}}%
\def\group#1{\refstepcounter{equation}\setcounter{saveeqn}
 {\value{equation}}%
  \label{#1}\setcounter{equation}{0}%
\renewcommand{\theequation}{\mbox{\arabic{section}.\arabic{saveeqn}
\alph{equation}}}
  \renewcommand{\)}{\end{equation}}}
\newcommand{\reseteqn}{\setcounter{equation}{\value{saveeqn}}%
  \renewcommand{\theequation}{\arabic{section}.\arabic{equation}}%
  \renewcommand{\)}{\end{equation}}}

\newcommand{\aalpheqn}{\setcounter{saveeqn}{\value{equation}}%
  \stepcounter{saveeqn}\setcounter{equation}{0}%
  \renewcommand{\theequation}{\mbox{
        \Alph{subsection}.\arabic{saveeqn}\alph{equation}}}
   \renewcommand{\)}{\end{equation}}}
\newcommand{\areseteqn}{\setcounter{equation}{\value{saveeqn}}%
  \renewcommand{\theequation}{\Alph{subsection}.\arabic{equation}}%
  \renewcommand{\)}{\end{equation}}}

\renewcommand{\thefootnote}{\alph{footnote}}
\renewcommand{\(}{\begin{equation}}
\renewcommand{\)}{\end{equation}}
\newcommand{\ba}{\begin{eqnarray}}
\newcommand{\ea}{\end{eqnarray}}
\newcommand{\cbp}{\mathop{\vtop{\ialign{##\crcr
   $\hfil\displaystyle{}\hfil$\crcr\noalign{\kern-13pt\nointerlineskip}
   \BIG{)}\hskip0pt\crcr\noalign{\kern3pt}}}}}
\newcommand{\pa}{\mathop{\vtop{\ialign{##\crcr

$\hfil\displaystyle{\oplus}\hfil$\crcr\noalign{\kern+1pt\nointerlineskip
}
   \hspace{.08in}$^{\alpha=0}$\hskip6pt\crcr\noalign{\kern3pt}}}}}
\renewcommand{\hsp}{,\hspace{.3in}}
\newcommand{\p}{^\prime}
\newcommand{\pp}{^{\prime\prime}}

\catcode`\@=11
\def\vereq#1#2{\lower3pt\vbox{\baselineskip1.5pt \lineskip1.5pt
\ialign{$\m@th#1\hfill##\hfil$\crcr#2\crcr\sim\crcr}}}
\catcode`\@=12

\renewcommand{\(}{\begin{equation}}
\renewcommand{\)}{\end{equation}}

\def\ppin#1{\int\hspace{-17pt}\sum \frac{d#1}{2\pi}}
\def\ppink#1{\int\hspace{-17pt}\sum\frac{d^{#1}k}{(2\pi)^{#1}}}

\def\dint{\int\hspace{-12pt}\sum }

\def\Ht{{\tilde H}}

\def\Bd#1{B^\ddag_{k_{#1}}}
\def\Bdp#1{B^\ddag_{k\p_{#1}}}

\def\df{\mathcal{D}_{f}}

\newcommand{\beas}{\begin{eqnarray*}}
\newcommand{\eeas}{\end{eqnarray*}}

\newcommand{\bquo}{\begin{quote}}
\newcommand{\enqu}{\end{quote}}

\def\lim#1{\stackrel{\rm{lim}}{{}_{#1}}}

\newcommand{\g}{\mathfrak g}

\def\ch{{\mathcal{H}}}

\def\ok#1{\omega_{k_{#1}}}

\newcommand{\beq}{\begin{equation}}
\newcommand{\eeq}{\end{equation}}
\newcommand{\bea}{\begin{eqnarray}}
\newcommand{\eea}{\end{eqnarray}}

\newskip\humongous \humongous=0pt plus 1000pt minus 1000pt

\newif\ifdtup

\catcode`\@=11

\ifthenelse{\equal{\letter}{0}}{ 

\@addtoreset{equation}{section}
\def\theequation{\arabic{section}.\arabic{equation}}

\def\@normalsize{\@setsize\normalsize{15pt}\xiipt\@xiipt
\abovedisplayskip 14pt plus3pt minus3pt%
\belowdisplayskip \abovedisplayskip
\abovedisplayshortskip \z@ plus3pt%
\belowdisplayshortskip 7pt plus3.5pt minus0pt}

\def\small{\@setsize\small{13.6pt}\xipt\@xipt
\abovedisplayskip 13pt plus3pt minus3pt%
\belowdisplayskip \abovedisplayskip
\abovedisplayshortskip \z@ plus3pt%
\belowdisplayshortskip 7pt plus3.5pt minus0pt
\def\@listi{\parsep 4.5pt plus 2pt minus 1pt
      \itemsep \parsep
      \topsep 9pt plus 3pt minus 3pt}}

\relax

\def\section{\@startsection{section}{1}{\z@}{3.5ex plus 1ex minus  .2ex}{2.3ex plus .2ex}{\large\bf}}

\def\thesection{\arabic{section}}
\def\thesubsection{\arabic{section}.\arabic{subsection}}

\def\appendix{\setcounter{section}{0}
 \def\thesection{Appendix \Alph{section}}
 \def\thesubsection{\Alph{section}.\arabic{subsection}}
 \def\theequation{\Alph{section}.\arabic{equation}}}
\renewcommand{\theequation}{\arabic{section}.\arabic{equation}}

}{
\renewcommand{\theequation}{\arabic{equation}}

} 

\begin{document}
\def\thefootnote{\fnsymbol{footnote}}
\def\thetitle{Quantum Oscillons May be Long-Lived}
\def\auttwo{Katarzyna Ole\'s}
\def\autone{Jarah Evslin}
\def\autthree{Tomasz Roma\'nczukiewicz}
\def\autfour{Andrzej Wereszczy\'nski}

\def\affc{Institute  of  Theoretical Physics,  Jagiellonian  University,  Lojasiewicza  11,  Krak\'ow,  Poland}
\def\affb{University of the Chinese Academy of Sciences, YuQuanLu 19A, Beijing 100049, China}
\def\affa{Institute of Modern Physics, NanChangLu 509, Lanzhou 730000, China}


\ifthenelse{\equal{\pr}{1}}{
\title{\thetitle}
\author{\autone}
\author{\auttwo}
\author{\autthree}
\affiliation {\affa}
\affiliation {\affb}

}{

\begin{center}
{\large {\bf \thetitle}}

\bigskip

\bigskip


{\large \noindent  \autone{${}^{1,2}$}\footnote{jarah@impcas.ac.cn} {
\autthree{${}^{3}$}\footnote{tomasz.romanczukiewicz@uj.edu.pl}
and \autfour{${}^{3}$}\footnote{andwereszczynski@gmail.com}
}}


\vskip.7cm

1) \affa\\
2) \affb\\
3) \affc\\

\end{center}

}

\begin{abstract}
\noindent
Hertzberg has constructed a quantum oscillon that decays into pairs of relativistic mesons with a power much greater than the radiation from classical oscillon decay.  This result is often construed as a proof that quantum oscillons decay quickly, and so are inconsequential.  We apply a construction similar to Hertzberg's to the quantum kink.  Again it leads to a rapid decay via the emission of relativistic mesons.  However, we find that this is the decay of a squeezed kink state to a stable kink state, and so it does not imply that the quantum kink is unstable.  We then consider a time-dependent solution, which may be an oscillon, and we see that the argument proceeds identically.  

\end{abstract}

%
\setcounter{footnote}{0}
\renewcommand{\thefootnote}{\arabic{footnote}}

\ifthenelse{\equal{\pr}{1}}
{
\maketitle
}{}

\section{Introduction}

Violent events in field theory, ranging from first order phase transitions to kink-antikink collisions, excite the available excitations.  After some time, only the longest lived excitations remain, and these dominate the dynamics.  A fairly generic result \cite{osc1,osc2,osc3} in classical field theory is that these longest lived excitations are oscillons \cite{sk}.  However, it is widely believed that in quantum field theory the oscillon lifetime is much shorter, and so they never dominate the dynamics, radically affecting the phenomenology of such events. 

This belief arises largely from two studies.  These considered a lift to quantum field theory of an oscillon \cite{hertzberg} and a breather \cite{tanmay} solution in classical field theory.  The authors found that the quantum state emits radiation, in particular pairs of relativistic mesons, at a much faster rate than the radiation emission in classical field theory.  Extrapolating the radiation power from the initial state to later times, one may conclude that the oscillon lifetime is much shorter in quantum field theory than in classical field theory.

These results are quite general and are derived robustly.  Ref.~\cite{hertzberg} uses standard \cite{riscalda} quantum field theory arguments in the Heisenberg picture, while Ref.~\cite{tanmay} uses the classical-quantum correspondence of Refs.~\cite{cqc1,cqc2}, which is reliable at order $O(g^0)$ in the coupling $g$.  Nonetheless, they both rely on a set of approximations.  Given the striking phenomenological consequences of this result, it is worth understanding the role played by these approximations, and their applicability to problems of interest.  

Most of the approximations are well-motivated.  For example, processes which affect the decay amplitude at $O(g)$ are ignored, and indeed these are subleading to the decays which are found.  However, both studies chose a specific lift of the classical field theory solution of interest.  In the case of Ref.~\cite{hertzberg} this choice is encoded in Eqs.~(30) and (31), which, as the author writes is ``its unperturbed vacuum state."  Similarly in Ref.~\cite{tanmay}, the initial condition is given in Eqs.~(17) and (18) which corresponds to the case in which, ``The quantum harmonic oscillators ... are taken to be in their ground state initially ... this corresponds to the quantum field $\psi$ being in its noninteracting vacuum."

Furthermore, it is difficult to connect the quantum evaporation of the sine-Gordon breathers \cite{tanmay} with the results obtained via the integrability \cite{dashen,Dorey, Zoltan}. Indeed, it is rigorously proven that, depending on the value of the coupling constant $g$, there are $n$ stable breathers at the quantum level, where this number is the integer part of $\sqrt{1-g^2}/g$. All these states are bound states of the first breather which exists until $g=1/\sqrt{2}$. In the semiclassical limit, where $g\to 0$, the number of stable breathers tends to infinity. However, one should be aware that these states, being Hamiltonian eigenstates, do not approach the Hertzberg states in the semiclassical limit. Nonetheless the existence of stable quantum breather states gives a hope that there is quantum lift which may lead to stable semiclassical breather and a long living quantum oscillon. 

The goal of the present note is to investigate how this choice of lift affects the decay of a quantum state into pairs of relativistic mesons.  If indeed the choice of lift affects the decay rate, it is important to understand how this choice may affect the oscillon decay rate reported in Ref.~\cite{hertzberg}.

We begin in Sec.~\ref{kinksez} by constructing a quantum state corresponding to a classical kink.  It is not the usual lift of Ref.~\cite{dhn2}, but rather is constructed by shifting the noninteracting vacuum.  We see that its one-loop energy is higher than that of the kink ground state and it decays into pairs of relativistic mesons, just like the quantum oscillon of Ref.~\cite{hertzberg}.  Needless to say, the end point of this decay is a lower energy, stable quantum kink, and so the strong initial radiation did not imply that the kink is unstable.  Next, in Sec.~\ref{oscsez}, we generalize this discussion to time-dependent solutions.  Now we no longer know whether there is a stable quantum ground state corresponding to the classical solution.  However, again we observe the same rapid decays into pairs of relativistic mesons.

\section{A Squeezed Kink State} \label{kinksez}
\subsection{The Classical Theory}

Let us consider a (1+1)-dimensional theory with a scalar field $\phi(x)$ and its conjugate momentum $\pi(x)$.  The classical Hamiltonian 
\beq
H=\int dx \ch(\phi(x),\pi(x))\hsp \ch(\phi(x),\pi(x))=\frac{\pi^2(x)+\partial_x\phi(x)\partial_x\phi(x)}{2}+\frac{V(g\phi(x))}{g^2} \label{hc}
\eeq
is characterized by a coupling constant $g$ and a potential $V$ with two degenerate minima.  Consider a stationary kink solution $\phi(x,t)=f(x)$ of the classical equations of motion that interpolates between these two minima.  It has a classical mass which can be written, for example, as
\beq
Q_0=\int dx\left[\frac{f^{\prime 2}(x)}{2}+\frac{V(gf(x))}{g^2}\right].
\eeq

We define the meson mass $m$ by
\beq
m=\sqrt{V\pp(gf(\infty))}=\sqrt{V\pp(gf(-\infty))}.
\eeq
We only are interested in potentials such that these two limits agree, as otherwise one-loop corrections will break the vacuum degeneracy and the kink will be accelerated from the false vacuum towards the true vacuum \cite{wstabile}.

\subsection{Plane Wave Expansion}

We pass to the quantum theory by imposing the canonical commutation relations
\beq
[\phi(x),\pi(y)]=i\delta(x-y) \label{ccr}
\eeq
on the Schrodinger picture operators $\phi(x)$ and $\pi(x)$.  We will work exclusively in the Schrodinger picture.   

We expand the field and its conjugate momentum in the plane wave basis
\beq
\phi(x)=\int \frac{dp}{2\pi} \left[
A^\ddag_p+\frac{1}{2\omega_p} A_{-p}
\right]
e^{-ipx}, \ \  \pi(x)=i\int \frac{dp}{2\pi} \left[ \omega_p A^\ddag_p- \frac{A_{-p}}{2 }\right] e^{-ipx},\ \  \omega_p=\sqrt{m^2+p^2}.
\eeq
This expansion is easily inverted
\beq
A^\ddag_p=\int dx e^{ipx}\left[ \frac{\phi(x)}{2}-i \frac{\pi(x)}{2\omega_p}\right]\hsp A_{-p}=\int dx e^{ipx}\left[ {\omega_p \phi(x)}+i {\pi(x)}\right]\hsp \label{aexp}
\eeq
which, using (\ref{ccr}), implies the operators $A^\ddag$ and $A$ are creation and annihilation operators
\beq
[A_p,A^\ddag_q]=2\pi \delta(p-q).
\eeq
Note that $\phi(x)$ and $\pi(x)$ are Hermitian, and so
\beq
A^\ddag_p=\frac{A^\dag_p}{2\omega_p}.
\eeq

Now that our operators do not commute, we need an ordering prescription to define our Hamiltonian $H$.  Thus we define the quantum Hamiltonian to be
\beq
H=\int dx :\ch(\phi(x),\pi(x)): \label{hq}
\eeq
where $::$ is the usual normal ordering that places all $A^\ddag_p$ to the left of $A_p$.

\subsection{The Hertzberg State}
Let us define the perturbative vacuum $|H\rangle$ to be that which satisfies
\beq
A_p|H\rangle=0
\eeq
for all $p$.  Note that
\beq
\langle H|A_p|H\rangle=\langle H|A^\ddag_p|H\rangle=0
\eeq
and so
\beq
\langle H|\phi(x)|H\rangle=0.
\eeq
We will define the Hertzberg kink state to be
\beq
|K_H\rangle=\df |H\rangle
\eeq
where the displacement operator $\df$ is defined to be
\beq
\df={{\rm Exp}}\left[-i\int dx f(x)\pi(x)\right]
\eeq
and $f(x)$ is a stationary kink solution of the classical equations of motion for $\phi(x,t)$.  This is called a displacement operator because
\beq
\df^\dag \phi(x)\df=\phi(x)+f(x).
\eeq

The expectation value of the $\phi(x)$ field on this state is
\beq
\frac{\langle H|\df^\dag \phi(x)\df |H\rangle}{\langle H|H\rangle}=\frac{\langle H|\phi(x)|H\rangle}{\langle H|H\rangle}+f(x)=f(x).
\eeq
Thus the Hertzberg kink state indeed describes a kink.

\subsection{Time Evolution}

Let the Hertzberg state evolve to a time $t$ and call it $|K_H(t)\rangle$
\beq
|K_H(t)\rangle=\df |H(t)\rangle \hsp |K_H(0)\rangle=|K_H\rangle\hsp
|H(0)\rangle=|H\rangle \label{kh}
\eeq
where the first equality defines $|H(t)\rangle$.
Then
\beq
\partial_t |K_H(t)\rangle=-iH|K_H(t)\rangle=-i H\df |H(t)\rangle.
\eeq
Act on both sides with $\df^\dag$, using
\beq
[\df^\dag,\partial_t] =0
\eeq
to obtain
\beq
\partial_t |H(t)\rangle=-i \df^\dag H\df |H(t)\rangle=-i H\p |H(t)\rangle \label{eveq}
\eeq
where we have defined the kink Hamiltonian \cite{mekink}
\beq
H\p= \df^\dag H\df.
\eeq

This situation may be interpreted using the language of passive transformations, as the quantum version of the passive transformation $\phi(x,t)\rightarrow \phi(x,t)-f(x)$ in classical field theory\footnote{This transformation takes small perturbations about a kink to small perturbations about zero.}.  The state itself is $|K_H(t)\rangle$, and like all Schrodinger picture states it evolves via the action of the Hamiltonian $H$.  However, as a result of the definition in the first equality of Eq.~(\ref{kh}), it is completely characterized by another ket, $|H(t)\rangle$, which evolves via the action of the kink Hamiltonian $H\p$.  Thus the same kink state can be equivalently described using two different kets, $|K_H(t)\rangle$ and $|H(t)\rangle=\df^\dag|K_H(t)\rangle$, but if one uses the later than one must conjugate all operators by $\df$.  

More abstractly, $|\psi\rangle\rightarrow \df^\dag |\psi\rangle$ is a passive transformation, changing the ket as a vector in the Hilbert space but still representing the same state.  Like all passive transformations, it must be accompanied by a transformation of all functions acting on the state, such as the Hamiltonian.  We refer to the $|K_H(t)\rangle$ representation of the state as the defining frame, and the $|H(t)\rangle$ representation as the kink frame.  This passive transformation viewpoint is intuitive, but is not essential for any calculations below.

Using Eq.~(\ref{hq}), one can show \cite{mekink} that
\beq
H\p=\int dx :\ch(\phi(x)+f(x),\pi(x)):. \label{kh2}
\eeq
Decomposing
\beq
H\p=\sum_{n=0}^\infty H\p_n
\eeq
where $H\p_n$ contains $n$ powers of fields when normal ordered, one finds that
\beq
H\p_0=\int dx \left[\frac{f\p(x)^2}{2}+\frac{V(gf(x))}{g^2}\right]=Q_0
\eeq
is the classical energy of the kink.  $H\p_1$ vanishes as $f(x)$ solves the classical equations of motion.  The terms
\beq
H\p_{n\geq 3}=g^{n-2}\int dx \frac{V^{(n)}(gf(x))}{n!}:\phi^n(x):
\eeq
are all suppressed by powers of the coupling $g$.  Thus the only operator which is not a $c$-number and not suppressed by powers of $g$ is $H\p_2$.

\subsection{Energy of the Hertzberg State}
Even though we have not yet evaluated $H\p_2$, we already have enough information to evaluate the energy of the Hertzberg state.  It is
\beq
\frac{\langle H|\df^\dag H\df|H\rangle}{\langle H|\df^\dag \df|H\rangle}=\frac{\langle H|H\p |H\rangle}{\langle H|H\rangle}=\frac{\langle H|H\p_0 |H\rangle}{\langle H|H\rangle}=Q_0.
\eeq
The key step, the second equality, follows because each $H\p_n$ is normal ordered, and so has all $A$ on the right, which annihilate $|H\rangle$ and $A^\dag$ on the left, which annihilate $\langle H|$.  The only exception is $H\p_0$ which is a $c$-number and contains neither $A$ nor $A^\dag$.  Thus the quantum energy of the Hertzberg state exactly equals its classical energy $Q_0$.  In contrast, it has long been known \cite{dhn2,cahill76} that the kink ground state has a negative quantum correction to its mass, of order $O(m)$.  Thus the Hertzberg state $\df|H\rangle$ has more energy than the kink ground state.  If it were an eigenstate of $H$ it would nonetheless be stable.  We will now show that it is not an eigenstate, and so is not stable.

\subsection{The Kink Hamiltonian in the Plane Wave Expansion}

We have argued that the evolution of a kink state is dominated by the action of $H\p_2$, which consists of all terms in Eq.~(\ref{kh2}) that are bilinear in the fields.   Substituting Eq.~(\ref{hc}) into Eq.~(\ref{kh2}) this is
\beq
H\p_2
=\int dx\left(  \frac{1}{2} :\pi^2: + \frac{1}{2} :(\partial_x\phi)^2: + \frac{1}{2}V''(gf) :\phi^2:\right). \label{hp2}
\eeq
We begin with the first term
\bea
\int dx   :\pi^2: &=& \int \int \frac{dp}{2\pi}\frac{dq}{2\pi} \int dx e^{-ix(p+q)} (-\omega_p\omega_q) :  \left( A^\ddag_p- \frac{1}{2\omega_p }A_{-p}\right) \left( A^\ddag_q- \frac{1}{2\omega_q }A_{-q}\right): \nonumber \\
&=&- \int \int \frac{dp}{2\pi}\frac{dq}{2\pi} 2\pi \delta(p+q) \omega_p\omega_q \left(  A^\ddag_pA^\ddag_q - \frac{1}{2\omega_p }A^\ddag_q A_{-p} - \frac{1}{2\omega_q }A^\ddag_p A_{-q}+  \frac{1}{4\omega_p\omega_q } A_{-p}A_{-q}\right) \nonumber \\
&=&- \int  \frac{dp}{2\pi}  \omega_p^2\left( A^\ddag_p A^\ddag_{-p}  - \frac{1}{2\omega_p }A^\ddag_{-p} A_{-p} - \frac{1}{2\omega_p }A^\ddag_p A_{p}+  \frac{1}{4\omega_p^2 } A_{-p}A_{p}\right) \nonumber \\
&=&-  \int \frac{dp}{2\pi}  \omega_p^2\left( A^\ddag_p A^\ddag_{-p}  - \frac{1}{\omega_p }A^\ddag_{p} A_{p}+ \frac{1}{4\omega_p^2 } A_{-p}A_{p}\right).
\eea
The second term is
\bea
\int dx  :\phi^{\prime 2}: &=& \int \int \frac{dp}{2\pi}\frac{dq}{2\pi} \int dx e^{-ix(p+q)} (-pq) :  \left( A^\ddag_p+ \frac{1}{2\omega_p }A_{-p}\right) \left( A^\ddag_q+ \frac{1}{2\omega_q }A_{-q}\right): \nonumber \\
&=&- \int \int \frac{dp}{2\pi}\frac{dq}{2\pi} 2\pi \delta(p+q) pq \left(  A^\ddag_pA^\ddag_q + \frac{1}{2\omega_p }A^\ddag_q A_{-p} +\frac{1}{2\omega_q }A^\ddag_p A_{-q}+  \frac{1}{4\omega_p\omega_q } A_{-p}A_{-q}\right) \nonumber \\
&=& \int  \frac{dp}{2\pi}  p^2\left( A^\ddag_p A^\ddag_{-p}  + \frac{1}{2\omega_p }A^\ddag_{-p} A_{-p} + \frac{1}{2\omega_p }A^\ddag_p A_{p}+  \frac{1}{4\omega_p^2 } A_{-p}A_{p}\right) \nonumber \\
&=&  \int \frac{dp}{2\pi}  p^2\left( A^\ddag_p A^\ddag_{-p}  + \frac{1}{\omega_p }A^\ddag_{p} A_{p}+ \frac{1}{4\omega_p^2 } A_{-p}A_{p}\right).
\eea
The third term
\bea
\int dx    V'':\phi^2: &=& \int \int \frac{dp}{2\pi}\frac{dq}{2\pi} \int dx e^{-ix(p+q)} V'':  \left( A^\ddag_p+ \frac{1}{2\omega_p }A_{-p}\right) \left( A^\ddag_q+ \frac{1}{2\omega_q }A_{-q}\right): \nonumber \\
&=& \int \int \frac{dp}{2\pi}\frac{dq}{2\pi} \int dx e^{-ix(p+q)} V'' \left(  A^\ddag_pA^\ddag_q + \frac{1}{2\omega_p }A^\ddag_q A_{-p} +\frac{1}{2\omega_q }A^\ddag_p A_{-q}+  \frac{1}{4\omega_p\omega_q } A_{-p}A_{-q}\right) \nonumber \\
&=& \int \int \frac{dp}{2\pi}\frac{dq}{2\pi}\tilde{V}''_{p+q} \left(  A^\ddag_pA^\ddag_q + \frac{1}{2\omega_p }A^\ddag_q A_{-p} +\frac{1}{2\omega_q }A^\ddag_p A_{-q}+  \frac{1}{4\omega_p\omega_q } A_{-p}A_{-q}\right).
\eea

So, if we act with this Hamiltonian on the perturbative vacuum $|H\rangle$, defined by $A_p|H\rangle=0$ we get
\beq
H'_2|H\rangle=\frac{1}{2} \int \frac{dp}{2\pi} (-\omega_p^2 +p^2 ) A^\ddag_p A^\ddag_{-p}  |H\rangle  + \frac{1}{2} \int \int \frac{dp}{2\pi}  \frac{dq}{2\pi} \tilde{V}''_{p+q}  A^\ddag_p A^\ddag_{q} |H\rangle. \label{paii} 
\eeq
Using (\ref{eveq}) this implies that, up to corrections of order $O(g)$
\beq
\partial_t|H(t)\rangle|_{t=0}=-iH\p|H\rangle=-i(Q_0+H\p_2)|H\rangle
\eeq
is not proportional to $|H\rangle$.  Instead pairs of mesons are created, corresponding to $A^\dag A^\dag|H\rangle$ in Eq.~(\ref{paii}).  Of these, some, before clearing the kink, will be reabsorbed via the $A^2$ term in $H\p_2$.  To avoid this complication, and so learn whether radiation escapes to infinity, we should work in the normal mode basis.

In conclusion, acting with the Hamiltonian on the Hertzberg state we produce pairs of mesons i.e., radiation. So, we could (wrongly) conclude that the quantum kink is unstable. But this is only because we chose a wrong ground state.  The correct ground state is  $\vac$ which is annihilated by $B$ operators, which are annihilation operators of the discrete and continuum normal modes \cite{cahill76}. 

\subsection{The Normal Mode Expansion}

With this motivation, following Refs.~\cite{cahill76,mekink}, we consider the expansion of the field and its conjugate momentum in the normal mode basis\footnote{We remind the reader that we are working in the Schrodinger picture, where the fields only depend on position.  Therefore they may be decomposed in any basis of the space of functions of one variable.  The normal modes provide such a basis as they solve a Sturm-Liouville equation.}
\beq
\phi(x)=\phi_0 \g_B(x) +\ppin{k} \g_k(x)\left(B^\ddag_k+\frac{B_{-k}}{2\ok{}}\right)\hsp
\pi(x)=\pi_0 \g_B(x)+ i\ppin{k} \g_k(x)\left(\ok{} B^\ddag_k-\frac{B_{-k}}{2}\right).\label{bexp}
\eeq
Here the normal modes $\g(x)$ correspond to solutions of the linearized classical equations of motion of the form $f(x)+\g(x)e^{-i\omega t}$.  There are three kinds of normal modes, classified by the value of the frequency $\omega$.  First there is the zero-mode $\g_B(x)$ with frequency $\omega_B=0$.  Next, there are continuum modes $\g_k(x)$ with $\omega_k=\sqrt{m^2+k^2}$.  Finally, some kinks have discrete, real shape modes $\g_S(x)$ with $0<\omega_S<m$.  The symbol $\dint dk/(2\pi)$ represents an integral $\int dk/(2\pi)$ over continuum modes plus a sum $\sum_S$ over shape modes.  They are normalized so that discrete modes integrate to unity and continuum modes integrate to Dirac delta functions.

As the normal modes are a basis, the transformation (\ref{bexp}) is invertible.  Therefore the canonical commutation relations (\ref{ccr}) lead to the commutation relations of the normal mode operators
\beq
[B_{k_1},\Bd2]=2\pi\delta(k_1-k_2)\hsp
[B_{S_1},B^\ddag_{S_2}]=\delta_{S_1,S_2}\hsp
[\phi_0,\pi_0]=i.
\eeq
Thus, $B^\ddag$ and $B$ are creation and annihilation operators for continuum and shape modes.  On the other hand, $\phi_0/\sqrt{Q_0}$ and $\sqrt{Q_0}\pi_0$ are the position and momentum operators of the kink center of mass.

Inserting (\ref{bexp}) into (\ref{aexp}) we arrive at the Bogoliubov transformation
\bea
A_{-p}&=&\frac{1}{2}\int dx e^{ipx}\left(\phi(x)+i\frac{\pi(x)}{\omega_p}\right)\label{abog}\\
&=&\frac{\tilde\g_B(-p)}{2}\left(\phi_0+i\frac{\pi_0}{\omega_p}\right)
+\ppin{k}\frac{\tilde\g_k(-p)}{\omega_p}\left[ 
\left(\omega_p-{\ok{}}{}
\right)B^\ddag_k
+\left(\omega_p+\omega_k\right)\frac{B_{-k}}{2\ok{}}
\right]
\nonumber\\
&=&M_{pk}B_{-k}+M_{pB}\phi_0+\hat{M}_{pk}\Bd{}+\hat{M}_{pB}\pi_0\nonumber
\eea
where $\tilde\g$ is the Fourier transform of $\g$.  In the last line we have introduced an abstract matrix notation, defined by the previous line, so that
\bea
M_{pk}&=&(\omega_p+\ok{})\frac{\tilde\g_k(-p)}{2\omega_p\ok{}}\hsp M_{pB}=\frac{\tilde\g_B(-p)}{2}\\
\hat M_{pk}&=&(\omega_p-\ok{})\frac{\tilde\g_k(-p)}{\omega_p}\hsp \hat M_{pB}=\frac{\tilde\g_B(-p)}{2\omega_p}.\nonumber
\eea
In particular, the contraction of two $p$ indices represents integration over $p$ with a factor of $1/2\pi$ while the contraction of two $k$ indices represents an integral over continuum modes with a factor of $1/2\pi$ plus a sum over discrete nonzero modes.  The $B$ index appears on the right side of the $M$ matrix, and its contraction is just ordinary multiplication.  

The formal inverse of the matrix $M$ then satisfies
\beq
\left(M^{-1}\right)_{kp}M_{pk\p}=2\pi\delta(k-k\p)\hsp
\left(M^{-1}\right)_{kp}M_{pB}=0. \label{inv}
\eeq
Note that if $k$ is a discrete index, representing a shape mode, then $2\pi\delta(p-p\p)$ should be replaced with $\delta_{pp\p}$.  This follows from the fact that $M^{-1}M$ is the identity matrix.  

Acting $M^{-1}$ on Eq.~(\ref{abog}) one obtains
\beq
B_{-k}=\left(M^{-1}\right)_{kp}\left[A_{-p}- \hat{M}_{pk}\Bd{}-\hat{M}_{pB}\pi_0
\right]. \label{bbog}
\eeq

\subsection{Radiation from the Hertzberg State}
In this basis, the $O(g^0)$ part of the kink Hamiltonian is just \cite{cahill76,mekink}
\beq
H\p_2=Q_1+\frac{\pi_0^2}{2}+\ppin{k}\ok{}\Bd{}B_k. \label{h2c}
\eeq
Here $Q_1$ is negative and is the one-loop correction to the kink mass.  The ground state of this leading order Hamiltonian is the leading order kink ground state $\vac$, defined by
\beq
\pi_0\vac=B_k\vac=0.
\eeq
Physically, $\pi_0\vac=0$ implies that the kink has no momentum, while $B_k\vac=0$ implies that the normal modes are in their ground states.

The state $\vac$ is the perturbative ground state of the one-kink sector.  Therefore a basis of the one-kink sector, in the kink frame, consists of the state $\vac$ acted upon with any number $m$ of zero modes and $n$ meson creation operators
\beq
\phi_0^m\Bd1\cdots\Bd n\vac.
\eeq
We may expand $|H\rangle$ in this basis
\beq
|H\rangle=\sum_{m,n} \ppink{n}\gamma^{mn}(k_1\cdots k_n) \phi_0^m\Bd1\cdots\Bd n\vac. 
\eeq
In other words, $|H\rangle$ is not the kink ground state, it is a superposition of the ground state with various excited states.  

We can be a bit more precise.  Recall that $|H\rangle$ is annihilated by $A_p$, while $\vac$ is annihilated by $\pi_0$ and $B_k$.  Recall also that $\pi_0$ and $B_k$ are related to $A_p$ by a Bogoliubov transformation.  We conclude that $|H\rangle$ is the squeezed state constructed by acting on $\vac$ with the squeeze operator that performs this Bogoliubov transformation.  Thus the decay of $|H\rangle$ is the usual decay of a squeezed state, and the fact that it decays into pairs of mesons results from the fact that the squeeze operator is the exponential of a bilinear in the creation operators.

We will be interested in the ground state component of $|H\rangle$
\beq
\gamma^{00}=\frac{\langle 0|H\rangle}{\langle 0\vac}
\eeq
and the 2-meson component
\beq
\gamma^{02}(k_1,k_2)=\frac{\langle 0|B_{k_1}B_{k_2}|H\rangle}{2\langle 0\vac}.
\eeq
With this motivation, we will calculate the following quantity, which is independent of the normalizations of $\vac$ and $|H\rangle$
\beq
2\frac{\gamma^{02}(-k_1,-k_2)}{\gamma^{00}}=\frac{\langle 0|B_{-k_1}B_{-k_2}|H\rangle}{\langle 0|H\rangle}. \label{duemeson}
\eeq
This quantifies the fraction of the $|H\rangle$ state that contains two unbound mesons, which proceed immediately to escape.  

What have we gained by using normal modes instead of plane waves?  Now our evolution operator (\ref{h2c}) contains only $\Bd{}B_k$, no terms which change the number of normal modes or even their frequencies.  Therefore any normal modes present in the initial state $|H\rangle$ will persist.  If they are continuum modes, this means that they will escape to infinity.  Thus the quantity in Eq.~(\ref{duemeson}) characterizes how much two-meson radiation will escape to infinity.

The matrix element is easily evaluated using Eq.~(\ref{bbog})
\bea
\frac{\langle 0|B_{-k_1}B_{-k_2}|H\rangle}{\langle 0|H\rangle}&=&\left(M^{-1}\right)_{k_1p_1}\left(M^{-1}\right)_{k_2p_2}\label{melt}\\
&&\times\frac{\langle 0|\left[A_{-p_1}- \hat{M}_{p_1k\p_1}\Bdp{1}-\hat{M}_{p_1B}\pi_0
\right]\left[A_{-p_2}- \hat{M}_{p_2k\p_2}\Bdp{2}-\hat{M}_{p_2B}\pi_0
\right]|H\rangle}{\langle 0|H\rangle}\nonumber\\
&=&-\left(M^{-1}\right)_{k_1p_1}\left(M^{-1}\right)_{k\p_2p_2}\frac{\langle 0|A_{-p_1}\left(\hat{M}_{p_2k\p_2}\Bdp{2}+\hat{M}_{p_2B}\pi_0
\right)|H\rangle}{\langle 0|H\rangle}\nonumber\\
&=&-\left(M^{-1}\right)_{k_1p_1}\left(M^{-1}\right)_{k_2p_2}\frac{\langle 0|\left(\hat{M}_{p_2k\p_2}[A_{-p_1},\Bdp{2}]+\hat{M}_{p_2B}[A_{-p_1},\pi_0]
\right)|H\rangle}{\langle 0|H\rangle}.\nonumber
\eea
Again using (\ref{abog}) one easily evaluates the commutators
\beq
[A_{-p_1},\Bdp{2}]=M_{p_1 k\p}[B_{k\p},\Bdp{2}]=M_{p_1 k\p_2}
\hsp
[A_{-p_1},\pi_0]=M_{p_1 B}[\phi_0,\pi_0]=iM_{p_1 B}.
\eeq
These two matrices can be contracted with the $\left(M^{-1}\right)_{k_1p_1}$ in Eq.~(\ref{melt}) using (\ref{inv}), to yield
\bea
\frac{\langle 0|B_{-k_1}B_{-k_2}|H\rangle}{\langle 0|H\rangle}&=&-\left(M^{-1}\right)_{k_1p_1}\left(M^{-1}\right)_{k_2p_2}\frac{\langle 0|\left(\hat{M}_{p_2k\p_2}M_{p_1 k\p_2}+i\hat{M}_{p_2B}M_{p_1 B}
\right)|H\rangle}{\langle 0|H\rangle}\\
&=&-\left(M^{-1}\right)_{k_2p_2}\frac{\langle 0|\hat{M}_{p_2k_1}
|H\rangle}{\langle 0|H\rangle}=-\left(M^{-1}\right)_{k_2p_2}\hat{M}_{p_2k_1}.
\nonumber
\eea

This quantity is not quite the amplitude for the emission of two mesons, because of the peculiar $\langle 0|H\rangle$ in the denominator.  However, it is proportional to this amplitude, with a $k_1$ and $k_2$-independent constant of proportionality, and also it is independent of the normalization of $\vac$ and $|H\rangle$.  More precisely, this matrix element is the ratio of the 2-meson part $\Bd 1\Bd 2\vac$ of $|H\rangle$ to the zero meson part $\vac$ of $|H\rangle$.  The fact that it is nonzero illustrates that $|H\rangle$ is a superposition of states, at least one of which $\Bd 1\Bd 2\vac_0$ contains pairs of mesons.  These mesons travel away from the kink at relativistic speeds, and so escape in a time $1/m$.  One exception is the case where $k_1$ and $k_2$ are both shape modes.  The doubly-excited shape mode is unstable in models such as the $\phi^4$ double well, and decays with a lifetime \cite{alberto} of order $O(1/g^2m^2)$, which at weak coupling is much longer than the other modes.  Also, it may be that either $k_1$ or $k_2$ is a shape mode and the other is a continuum mode, in which case, after relaxing this summand in the state $|H\rangle$, one arrives at an excited kink.  We conclude that $|H\rangle$ decays to a superposition of the ground state kink with its stable excitations.

\subsection{The Inverse of $M$}

From Eq.~(2.13) of Ref.~\cite{me2loop}
\beq
\tilde\g_B(p)\tilde\g_B(-q)+\ppin{k}\tilde\g_k(p)\tilde\g_{-k}(-p)=2\pi\delta(p-q). \label{ginv}
\eeq
Ignoring the zero-mode, this means that the inverse of $\tilde \g_k(p)$ is $\tilde\g_{-k}(-p)$.  

But how do we invert $M$?  We first decompose
\beq
M_{pk}=A_{pk}+B_{pk}\hsp A_{pk}=\frac{\tilde\g_k(-p)}{\ok{}}\hsp B_{pk}=(\ok{}-\omega_p)\frac{\tilde\g_k(-p)}{2\ok{}\omega_p}
\eeq
while the zero-mode column is
\beq
M_{pB}=A_{pB}+B_{pB}
\hsp A_{pB}=\frac{\tilde\g_B(-p)}{2}\hsp B_{pB}=0.
\eeq
Eq.~(\ref{ginv}) implies
\beq
\left(A^{-1}\right)_{kp}=\omega_k\tilde\g_{-k}(p)\hsp
\left(A^{-1}\right)_{Bp}=2\tilde\g_B(p).
\eeq
The matrix $\tilde\g_k(p)$ is supported near $k\sim p$, with an exponential falloff in $(k-p)/m$.  In this sense, $A$ is much larger than $B$ for ultrarelativistic $k\gg m$.  This justifies the expansion
\beq
M^{-1}=(A+B)^{-1}=A^{-1}(1+BA^{-1})^{-1}=A^{-1} \sum_{n=0}^\infty (-BA^{-1})^n.
\eeq
Here
\bea
-\left(BA^{-1}\right)_{p_1 p_2}&=&-B_{p_1 B}\left( A^{-1}\right)_{B p_2}-B_{p_1 k}\left( A^{-1}\right)_{k p_2}\\
&=&\frac{1}{2\omega_{p_1}}\ppin{k} {(\omega_{p_1}-\ok{})}{}
\tilde\g_k(-p_1)\tilde\g_{-k}(p_2).\nonumber
\eea
This yields the usual geometric series sum of the propagator, where one recognizes $BA^{-1}$ as the Green's function defined in Ref.~\cite{gjs75}.

\section{Time-Dependent Solutions} \label{oscsez}

\subsection{The Defining Frame}

Again, we consider a (1+1)-dimensional field theory with operators $\phi(x)$ and $\pi(x)$ satisfying canonical commutation relations (\ref{ccr})
\beq
[\phi(x),\pi(y)]=i\delta(x-y).
\eeq
Define a theory using the Hamiltonian
\beq
H=\int dx :\ch(x):\hsp
\ch(x)=\frac{\pi^2(x)+(\partial_x \phi(x))^2}{2}+\frac{V(g\phi(x))}{g^2}.
\eeq
We no longer demand that $V$ have degenerate minima.  The classical Hamilton equations are
\beq
\dot \phi(x,t)=\frac{\delta H}{\delta \pi(x)}= \pi(x,t)\hsp
\ddot\phi(x,t)=\dot \pi(x,t)=-\frac{\delta H}{\delta \phi(x)}=\partial^2_x \phi(x,t)-\frac{V\p(g\phi(x,t))}{g}.
\eeq
Fix a classical solution $\phi(x,t)=f(x,t)$ to this equation
\beq
\ddot f=\partial^2_x f-\frac{V\p(gf)}{g}. \label{ffeq}
\eeq

Now consider a nearby solution $\phi(x,t)=f(x,t)+\g(x,t)$.  The classical equations of motion are
\beq
\ddot f+\ddot \g=\partial^2_x f+\partial^2_x \g-\frac{V\p(g(f+\g))}{g}.
\eeq
Subtracting Eq.~(\ref{ffeq}) yields
\beq
\ddot \g=\partial^2_x \g-\frac{V\p(g(f+\g))}{g}+\frac{V\p(g\g)}{g}=\partial^2_x \g-\sum_{n=1}^\infty \frac{g^{n-1}  V^{(n+1)}(gf)}{n!} \g^n
.
\eeq
At linear order in $\g$ this simplifies to
\beq
\ddot \g=\partial^2_x \g-V^{(2)}(gf) \g
.
\eeq

\subsection{The Comoving Frame}
Define the Schrodinger picture displacement operator
\beq
\df^{(t)}=\exp{i\int dx (\dot f(x,t) \phi(x)-f(x,t)\pi(x))}.
\eeq
Recall that the Schrodinger picture operators $\phi(x)$ and $\pi(x)$ are independent of $t$.  The displacement operator performs the passive transformation
\beq
\phi\p(x)=\df^{(t)\dag}\phi(x)\df^{(t)}=\phi(x)+f(x,t)\hsp
\pi\p(x)=\df^{(t)\dag}\pi(x)\df^{(t)}=\pi(x)+\dot f(x,t)
\eeq
which does depend on $t$, although we leave the $t$-dependence of $\phi\p(x)$ and $\pi\p(x)$ implicit so that these fields are not confused with Heisenberg picture fields.

Similarly we define
\beq
H^{\prime(t)}[\phi,\pi,t]=\df^{(t)\dag}H[\phi,\pi]\df^{(t)}=H[\phi+f,\pi+\dot f]
\eeq
which is unitarily equivalent to $H$ and so has the same spectrum.  It is easily evaluated
\bea
H^{\prime(t)}&=&\int dx :\ch^{\prime(t)}(\phi,\pi):\\
\ch^{\prime(t)}(\phi,\pi)&=&\frac{(\pi+\dot f)^2+(\partial_x (\phi+f))^2}{2}+\frac{V(g(\phi+f))}{g^2}.
\eea

We may again represent each state with two distinct kets, corresponding to two frames.  If a state at time $t$ is represented in the defining frame by the ket $|\psi(t)\rangle$, then we define a comoving frame in which the same state is denoted by the ket $\df^{(t)\dag}|\psi(t)\rangle$.  This ket is comoving in the sense that the expectation value of the field $\phi(x)$ is shifted by $-f(x)$
\beq
\frac{\langle \psi(t)|\df^{(t)} \phi(x)\df^{(t)\dag}|\psi(t)\rangle}{\langle \psi(t)|\df^{(t)}\df^{(t)\dag}|\psi(t)\rangle}=\frac{\langle \psi(t)| \phi(x)|\psi(t)\rangle}{\langle \psi(t)|\psi(t)\rangle}-f(x,t)
\eeq
and so it is comoving with the solution in field space.

How do comoving frame kets evolve?  As defining frame kets $|\psi(t)\rangle$ evolve via the action of $H$, one finds that comoving frame kets $\df^{(t)\dag}|\psi(t)\rangle$ evolve via
\bea
\partial_t\df^{(t)\dag}|\psi(t)\rangle&=&\df^{(t)\dag}\df^{(t)}\partial_t\df^{(t)\dag}|\psi(t)\rangle=\df^{(t)\dag}\partial_t|\psi(t)\rangle+\df^{(t)\dag}[\df^{(t)},\partial_t]\df^{(t)\dag}|\psi(t)\rangle\nonumber\\
&=&-i\df^{(t)\dag}H|\psi(t)\rangle+\df^{(t)\dag}[\df^{(t)},\partial_t]\df^{(t)\dag}|\psi(t)\rangle\nonumber\\
&=&-i\left(H\p-i\df^{(t)\dag}[\df^{(t)},\partial_t]\right)\df^{(t)\dag}|\psi(t)\rangle.
\eea
In other words, in the comoving frame, time translations are generated by
\beq
\Ht^{\prime(t)}=H^{\prime(t)}-i\df^{(t)\dag}[\partial_t,\df^{(t)}]. \label{tilde}
\eeq
The additional commutator term is the standard partial derivative term in the Hamiltonian in classical mechanics that arises when performing a canonical transformation with explicit time dependence.

It is interesting to remark that transition (\ref{tilde}) from the defining Hamiltonian $H$ to the comoving Hamiltonian $\tilde{H}^{'(t)}$ which generates time translation resembles the covariant derivative.

The commutator is easily calculated 
\bea
[\partial_t,\df^{(t)}]&=&\sum_{n=0}^\infty \frac{i^n}{n!}\left[\partial_t,\left(\int dx \phi(x) \dot f(x,t)-\pi(x) f(x,t)\right)^n\right]\\
&=&\sum_{n=0}^\infty \frac{i^n}{n!}\sum_{m=0}^{n-1}\left(\int dx \phi \dot f-\pi f\right)^{n-m-1}\left(\int dx \phi \ddot f-\pi \dot f\right)\left(\int dx \phi \dot f-\pi f\right)^{m}\nonumber\\
&=&\sum_{n=0}^\infty \frac{i^n}{n!}\sum_{m=0}^{n-1}
\left[\left(\int dx \phi \dot f-\pi f\right)^{n-1}\left(\int dx \phi \ddot f-\pi \dot f\right)\right.\nonumber\\
&&\left.+\left(\int dx \phi \dot f-\pi f\right)^{n-m-1}\left[\left(\int dx \phi \ddot f-\pi \dot f\right),\left(\int dx \phi \dot f-\pi f\right)^{m}\right]
\right]\nonumber
\\
&=&\sum_{n=0}^\infty \frac{i^n}{n!}\left[n\left(\int dx \phi \dot f-\pi f\right)^{n-1}\left(\int dx \phi \ddot f-\pi \dot f\right)\right.\nonumber\\
&&\left.+ \sum_{m=0}^{n-1}\left(\int dx \phi \dot f-\pi f\right)^{n-2}m\left(\int dx -if\ddot f+i\dot f^2\right)
\right]
\nonumber
\\
&=&i\df^{(t)}\int dx \left[\phi(x) \ddot f(x,t)-\pi(x) \dot f(x,t)+\frac{f\ddot(x,t) f(x,t)-\dot f^2(x,t)}{2}
\right]. 
\eea

Thus we conclude that the time translation operator in the comoving frame is the comoving Hamiltonian
\beq
\Ht^{\prime(t)}=H^{\prime(t)}+\int dx\left[\phi(x) \ddot f(x,t)-\pi(x) \dot f(x,t)+\frac{f\ddot(x,t) f(x,t)-\dot f^2(x,t)}{2}
\right].
\eeq
Let us decompose this into the terms $\Ht^{\prime(t)}_n$ in $\Ht^{\prime(t)}$ with $n$ powers of $\phi(x)$ and $\pi(x)$.  Now $\Ht^{\prime(t)}_0$ is arbitrary, as it may be modified by adding a phase to the displacement operator.  The first interesting term is the tadpole
\beq
\Ht^{\prime(t)}_1=H^{\prime(t)}_1+\int dx\left[\phi \ddot f-\pi \dot f\right]=\int dx 
\phi(x)\left(-\partial^2_x f+\frac{V\p(gf)}{g}+\ddot f
\right)=0
\eeq
where the last equality follows from Eq.~(\ref{ffeq}).  At all higher $n$, $\Ht\p$ and $H\p$ agree, and so the comoving Hamiltonian formally has the same form as the kink Hamiltonian.  For example
\beq
\Ht^{\prime(t)}_2=\frac{1}{2}\int dx \left(:\pi^2(x):+V\pp(gf):\phi^2(x):\right).
\eeq
We recognize this as the same time translation operator as in the case of the kink in Eq.~(\ref{hp2}).

Now we are done.  The calculation has been reduced to that of the kink.  One may define a Hertzberg state for any time-dependent solution, including an oscillon, as $\df^{(t)}|H\rangle$.  At any moment in time $t$ the state is $\df^{(t)}|H(t)\rangle$ where
\beq
\partial_t |H(t)\rangle=-i\Ht^{\prime(t)}|H(t)\rangle=-i\Ht^{\prime(t)}_2|H(t)\rangle+O(g).
\eeq
Now the calculation in the case of the kink shows that $\Ht^{\prime(t)}_2$ has terms of the form $A^\dag A^\dag$ which create pairs of relativistic mesons.  

To robustly show that these mesons escape to infinity requires an understanding of the normal modes, which is an open problem for unstable and time-dependent solutions such as the oscillon.  Nonetheless, it illustrates a counter example to the tempting logic that a large initial decay rate into pairs of relativistic mesons implies a short lifetime.  If a Bogoliubov transformation exists here that can transform the $A$ operators into at least the classically stable normal modes, then only the classically unstable ones will lead to an instability of the transformed state.  These unstable modes have wavelengths of order the oscillon size, and so correspond to highly nonrelativistic mesons, which are quite different from those of the fast decay process described by Hertzberg.  This lends hope to a conjecture that an oscillon state exists which decays much more slowly than the quantum oscillon of Ref.~\cite{hertzberg} or the breather of \cite{tanmay}.

\section{Summary and Speculations}

We have considered lifts of solutions of classical field theory to quantum states.  We have found that the decay rate of the quantum state to pairs of relativistic mesons depends not only on the classical solution, but also depends strongly on the choice of lift.  This implies that the determination of the quantum decay rate of a given classical solution should be phrased as a minimization problem, where one selects the quantum lift with the lowest decay rate.  This lift is the state that one expects to find after the oscillon has had time to relax.

In light of this result, it would be interesting to examine the dependence of the decay rates found in Refs.~\cite{hertzberg} and \cite{tanmay} on the choice of lift.  It is possible that, as in the case of the quantum kink, a lift of the oscillon may be found whose decay rate is qualitatively smaller than that of the lift considered in Ref.~\cite{hertzberg}. 

Wilder speculations are possible.  The oscillon is not the first system where a Bogoliubov transformation leads to an instability consisting of emission of pairs of particles.  In the case of an accelerating frame, such an argument leads to the Unruh effect \cite{unruh}.  When this frame corresponds to a timelike Killing vector beyond an event horizon, the corresponding radiation is called Hawking radiation \cite{hawking}.  

It is commonly believed that Hawking radiation continues until the black hole mass decreases either to zero, or at least until it becomes extremal.  This leads to numerous potential problems such as information loss \cite{hawkinginf}, naked singularities \cite{rn} and remnants \cite{rnussinov}.  Curing these problems has led to conjectures such as the weak gravity conjecture \cite{nimaweak}, the no global symmetry conjecture \cite{bsnoglob} and the completeness hypothesis \cite{complete}, which have become the pillars of modern thinking regarding quantum gravity.  However what if instead, like the Hertzberg kink, the black hole state described by Hawking only radiates a very small amount, of order the Planck scale perhaps, before settling into its proper one-loop ground state?  

\section* {Acknowledgement}

\noindent
JE is supported by NSFC MianShang grants 11875296 and 11675223 and the CAS Key Research Program of Frontier Sciences grant QYZDY-SSW-SLH006. TR and AW were supported by the Polish National Science Centre, 
grant NCN 2019/35/B/ST2/00059. AW thanks Zoltan Bajnok for discussion.

\end{document}